\documentclass[draft]{agujournal2019}
\usepackage{url} 
\usepackage{soul}
\draftfalse

\journalname{Geophysical Research Letters}

\begin{document}

%
%


\title{Reformation of Supercritical Perpendicular Shock}

%
%




\authors{Yuri V. Khotyaintsev\affil{1,2}, Domenico Trotta\affil{3}, Daniel B. Graham\affil{1} }

\affiliation{1}{Swedish Institute of Space Physics, Uppsala, Sweden}
\affiliation{2}{Space and Plasma Physics, Department of Physics and Astronomy, Uppsala University, Uppsala, Sweden}
\affiliation{3}{European Space Agency (ESA), European Space Astronomy Centre (ESAC), Camino Bajo del Castillo s/n, 28692 Villanueva de la Cañada, Madrid, Spain}




\correspondingauthor{Yuri Khotyaintsev}{yuri@irfu.se}



\begin{keypoints}
\item MMS observations and high-resolution simulation investigate the non-stationarity of a perpendicular supercritical collisionless shock.
\item The shock develops neighboring regions with a narrow Hall-dominated ramp and a broad reflected-ion foot.
\item Feedback between Hall-field ion reflection and the evolving ion foot explains several signatures of the observed shock non-stationarity.
\end{keypoints}

%
%

%
%


\begin{abstract}
Super-critical collisionless shocks are not static but evolve continuously as they reflect incoming ions upstream. The process responsible for this non-stationarity — whether wave-like corrugation of the shock surface (rippling) or cyclic rebuilding of the shock transition (reformation) — remains debated. We combine Magnetospheric Multiscale (MMS) observations of a nearly perpendicular ($\theta_{Bn}\approx89^\circ$), supercritical ($M_A\approx6$) bow shock with a high-resolution two-dimensional hybrid simulation. MMS reveals repeated ion phase-space holes and intense, localized Hall electric fields. A virtual-spacecraft analysis of the simulation reproduces these signatures and shows that they arise from a self-regulating feedback cycle: strong Hall-field ion reflection builds a reflected-ion foot, which weakens the Hall field and suppresses further reflection until the foot decays and the cycle restarts. This reformation cycle, spatially organized by the two-dimensional shock structure, can account for several principal signatures of the observed non-stationarity.
\end{abstract}

\section*{Plain Language Summary}
When the solar wind slams into Earth's magnetic field, it forms a shock wave, much like the sonic boom in front of a supersonic jet. This "bow shock" is not a smooth, unchanging wall of plasma — it constantly rebuilds itself as some of the incoming particles bounce off and pile up in front of it before the shock catches up and absorbs them. Exactly how this rebuilding happens has been debated for decades: does the shock's surface simply ripple like a flag in the wind, or does the entire shock front repeatedly break down and reform, like waves crashing on a beach? Using detailed measurements from NASA's Magnetospheric Multiscale (MMS) spacecraft, together with a computer simulation of a shock, we show that the second picture explains several key features, while others point to additional three-dimensional rippling. Strong electric fields at the shock repeatedly reflect incoming particles, building up a cushion of returning particles that temporarily weakens those fields, until the cushion drains away and the cycle begins again. This self-repeating cycle, occurring at slightly different times across the shock's surface, explains some of the key features of the disturbed, ever-changing structure seen by the spacecraft.

%
%

\section{Introduction}
Collisionless quasi-perpendicular supercritical shocks are intrinsically non-stationary, but the physical processes responsible for their variability remain a long-standing problem. Non-stationarity encompasses several phenomena, including cyclic shock reformation and spatial variations associated with shock-surface rippling. Early one-dimensional (1-D) particle simulations showed that periodic ion reflection can cause the shock ramp to decay and reform on ion timescales, a process known as shock reformation \cite{biskamp1972,Burgess1989,lembege1992}. 
Subsequent particle-in-cell (PIC) and hybrid simulations have extensively investigated this process and its dependence on shock parameters \cite{Hada2003,Scholer2003,Chapman2005,Lembege2009,Umeda2010,Yuan2009}. Although 1-D simulations provide a useful framework for understanding cyclic reformation, they exclude variations along the shock surface and therefore cannot capture the full range of multidimensional shock dynamics.

In two dimensions, shock reformation persists, though the dynamics become more complex as reflected ions gyrate upstream and return to the shock at locations displaced from their original reflection points. 2D geometry also permits the development of the Alfv\'en ion cyclotron (AIC) instability, which corrugates the shock front with ion-scale ripples \cite{winske1988}. These ripples locally rotate the shock normal and produce spatial variations in shock properties, including periodic changes in the normal magnetic field component, $B_n$~\cite{Lowe2003}. 
Fully 3D hybrid simulations further show that field-aligned AIC fluctuations can couple to the quasi-perpendicular reformation process, producing obliquely propagating ripples that combine characteristics of both surface corrugation and cyclic ion reflection \cite{burgess2016}. These results demonstrate that reformation and rippling are distinct but potentially coupled processes.
Previous one-dimensional studies have also documented extended ion feet and varying electric fields at different stages of the reformation cycle \cite{Lee2005POP,Lee2005AG,Yang2009POP,Yang2009JGR,Hao2023ApJ}. However, the relationship between these features and the spatially varying ion dynamics in two dimensions remains less well understood.

Spacecraft observations provide complementary evidence for non-stationary shock structure. Cluster measurements revealed shock-profile variations and ion-reflection signatures consistent with cyclic evolution at high-Mach-number quasi-perpendicular shocks \cite{Lobzin2007,Lefebvre2009,Mazelle2010}. MMS has shown that non-stationarity is common at Earth's bow shock: it resolved proton phase-space holes within the shock transition \cite{johlander2016,johlander2018}, demonstrated the importance of strong sub-proton-scale normal electric fields for ion reflection \cite{khotyaintsev2024_ion_reflection}, and found non-stationarity under a wide range of upstream conditions \cite{lotekar2025_occurrence}. Reformation-like behavior has also been identified in other geometries, including quasi-parallel shocks \cite{Johlander2022,Turner2021}. For some quasi-perpendicular crossings, the observations indicate formation of a new ramp \cite{Yang2020,Xu2026}. In other cases, however, the observations can be interpreted in terms of shock-surface rippling \cite{johlander2016,johlander2018,Madanian2021}, even though the reflected-ion population, electric field, and shock width still vary cyclically and no new ramp forms. Thus, it is not yet clear whether the observations are generally explained by obliquely propagating shock ripples, as suggested by three-dimensional simulations \cite{burgess2016}, or whether cyclic reformation physics dominates the observed non-stationarity.

Here we investigate a nearly perpendicular ($\theta_{Bn}\sim$ 89$^{\circ}$) bow shock observed by MMS, together with a high-resolution two-dimensional hybrid simulation. 
We investigate shock “reformation” in a broad sense, referring to the cyclic evolution of ion reflection and the related magnetic and electric field structure; this does not require a new ramp to form at a different location.
The simulation contains this physics but, with the background field out of the simulation plane, excludes AIC ripples. Using a virtual spacecraft, we show that reformation, coupled to the two-dimensional structure along the shock front, can account for several principal signatures of the observed non-stationarity, while the remaining differences point to three-dimensional effects.

\section{Observations}
\begin{figure}[!h]
 \center
\includegraphics[width=1\textwidth]{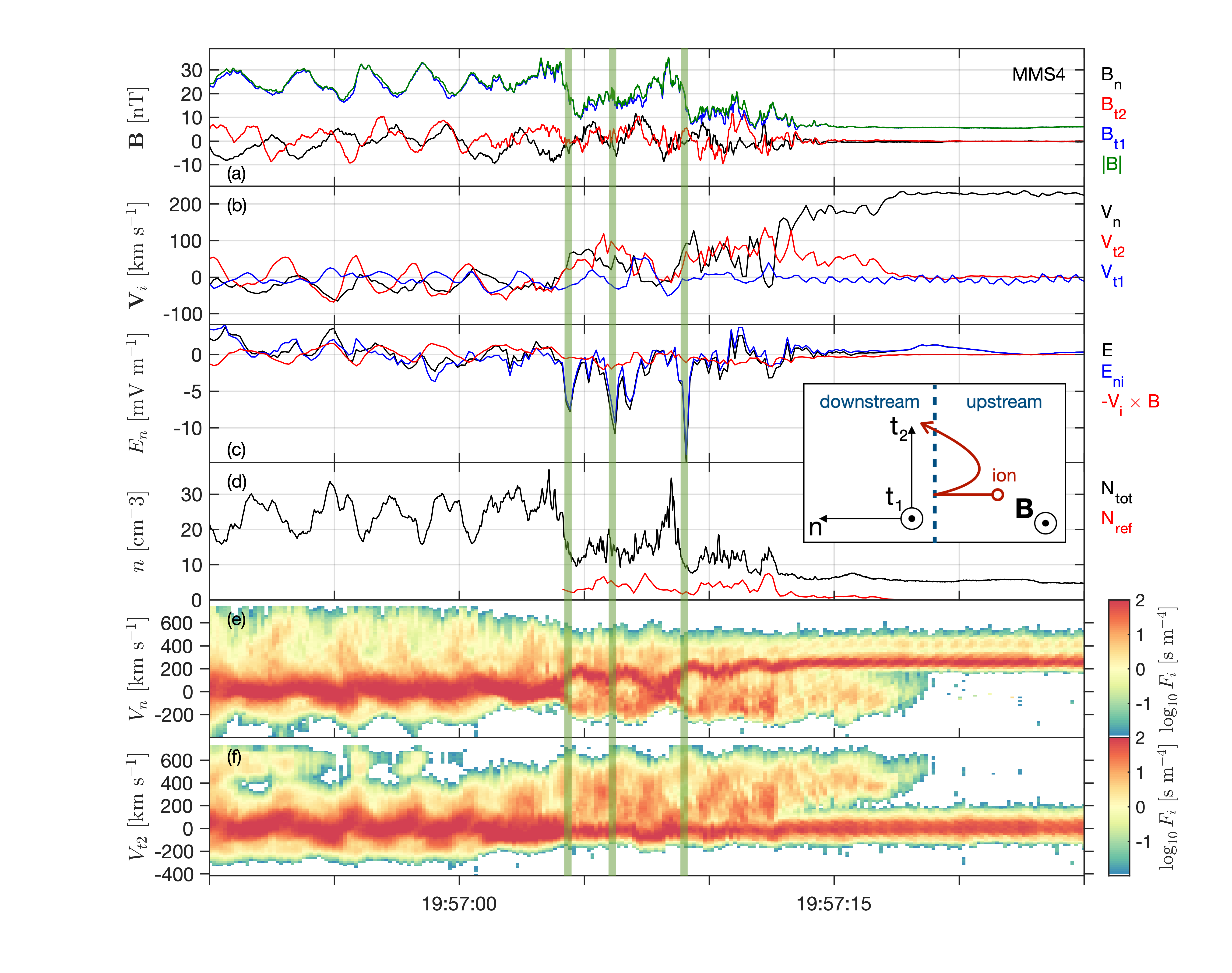}
\caption{Overview of the perpendicular shocks observed by MMS. Panels from top to bottom show: (a) magnetic field, (b) ion velocity, (c) normal component of the measured electric field (down-sampled to the cadence of ion measurements), non-ideal field $E_{ni}$, and ion convection, (d) total density in black and density of reflected ions in red, (e,f) reduced 1D ion velocity distribution functions (VDFs) as a function of $V_n$ and $V_{t2}$. All the data are using the $\mathbf{\hat{n}}$, $\mathbf{\hat{t}_2}$, $\mathbf{\hat{t}_1}$ coordinate system illustrated by the inset and are in the NIF downstream rest frame.}
\label{fig:mms}
\end{figure}

We analyze a nearly-perpendicular bow shock, $\theta_{Bn}\sim$ 89$^{\circ}$, observed by MMS on November 14, 2017, previously reported by \citeA{khotyaintsev2024_ion_reflection}. The shock parameters are summarized in Table~\ref{tab:params}. We use the magnetic field measured by FGM \cite{russell2014,mms4fgmbrst}, ion data by FPI-DIS \cite{pollock2016,mms4fpidisdistbrst,mms4fpidismomsbrst}, and the electric field measured by the electric double probes \cite{lindqvist2014, ergun2014,mms4sdpEbrst} on MMS4. Figure~\ref{fig:mms} shows an overview of the event. Throughout this letter we use a $\mathbf{\hat{n}}$, $\mathbf{\hat{t}_2}$, $\mathbf{\hat{t}_1}$ coordinate system, where $\mathbf{\hat{n}}$ is the shock normal, $\mathbf{\hat{t}_2} = \mathbf{\hat{B_u}} \times \mathbf{n}/|\mathbf{\hat{B_u}} \times \mathbf{n}|$ ($B_u$ is the upstream B) and $\mathbf{\hat{t}_1} = \mathbf{\hat{n}} \times \mathbf{\hat{t}_2}$ (see the inset) and the normal incidence (NIF) frame in which the downstream is stationary. We choose the normal pointing towards the downstream, which makes comparison with the simulation more convenient. The shock speed is $V_{sh} = [57, 0, 0 ]$ km/s in NIF. 

\begin{table}[b]
\caption{
Shock and upstream plasma parameters in the NIF shock rest frame.
}
\begin{tabular}{lc}
\hline
\textrm{Parameter}&
\textrm{Value}\\
\hline
Magnetic field magnitude $|B_u|$ & 6 nT\\
SW density $n_{sw}$ & 4.7  cm$^{-3}$\\
SW velocity $V_{sw}$ & 288 km s$^{-1}$\\
Alfv\'en Mach number $M_A$ & 6.4\\
Magnetosonic Mach number $M_{ms}$ & 5.3\\
SW ion $\beta_{i,u}$ & 0.2\\
Shock normal in GSE $\mathbf{\hat{n}}$ & [-0.92 -0.37 -0.10]\\
$\theta_{Bn}$ & 89$^{\circ }$\\
\hline

\label{tab:params}
\end{tabular}
\end{table}

The crossing is from the downstream to the upstream, so the shock is retreating towards the Earth. We observe a clear upstream after 19:57:20\,UT, $B$ is in the $\mathbf{\hat{t}_1}$ direction, $V_i$ is along $\mathbf{\hat{n}}$, consistent with NIF. The ion VDFs in panels (e) show a distinct solar wind proton peak at $V_n \sim 230$\,km/s. The secondary peak corresponds to alpha particles, see \citeA{khotyaintsev2024_ion_reflection}. The downstream is until 19:57:03\,UT, where we observe large-amplitude periodic oscillations in $|B|$ and density; The ion VDFs (panels e, f) show a thermalized proton core and a gyrating proton component having large $V_n$ and $V_{t2}$. 

The complex shock transition region is observed during the interval 19:57:03--19:57:20\,UT, containing large variations in B (e.g., two distinct steep ramps), density, velocity, and normal electric field, $E_n$. Ion VDFs are composed of the incoming slowed-down proton population and the reflected protons forming the foot,  having positive $V_{n}$ (panel e) and $V_{t2}$ (panel f). The presence of the foot is reflected in positive bulk $V_{t2}$ and reduced $V_n$ compared to the upstream (panel b). 

Reduced ion VDFs reveal multiple holes in ion phase space, Figure~\ref{fig:mms}e, indicating repeated disappearance and reappearance of the reflected-ion population. Such phase-space holes are a hallmark of a non-stationary shock, produced when motion of the ion reflection point repeatedly carries the spacecraft into and out of the foot \cite{johlander2016,lotekar2025_occurrence}. The density of reflected ions, calculated by integrating the ion VDF for $V_{t2} > 100$\,km/s, sometimes exceeds the upstream density, $n_{sw} \sim 4.7$\,cm$^{-3}$, consistent with the shock being non-stationary.

The shock transition contains several intense negative (the field pointing towards the upstream) spikes in the normal electric field $E_n$, Figure~\ref{fig:mms}c. These spikes substantially exceed the ion convection field and are primarily balanced by the non-ideal electric field $E_{ni}$, which is a measure of deviation of $E$ from $-\mathbf{V_i} \times \mathbf{B}$. According to the normal force balance at the shock using the generalized Ohm's law \cite{khotyaintsev2006}, neglecting electron inertia: 
\begin{equation}
\mathbf{E} + \mathbf{V}_i \times \mathbf{B} = \frac{1}{ne} \mathbf{J} \times \mathbf{B} 
- \frac{1}{ne} \nabla p_e = \mathbf{E}_{ni},
\label{eq:gol}
\end{equation}
both the Hall, $\mathbf{J} \times \mathbf{B} /ne$, and electron pressure gradient, $-\nabla p_e / ne$, terms contribute to the non-ideal electric field, $\mathbf{E}_{ni}$. The observed spikes in $E_n$ are mainly balanced by $E_{ni}$, and \citeA{khotyaintsev2024_ion_reflection} have shown that the dominant contribution is from the Hall term. We also note a small negative convection field, $-(\mathbf{V_i} \times \mathbf{B})_n <0$, correlated with increases in $V_{t2}$ and $N_{ref}$.

We mark the three most significant $E_n$ spikes by green lines in Figure~\ref{fig:mms}. Remarkably, all three strongest $E_n$ spikes occur when both $B$ and $n$ decrease across the ramp, indicating that the shock front is locally retreating downstream. They are also systematically located at the left edge of the ion phase-space holes, i.e., close to the ion reflection point (Figure~\ref{fig:mms}e). In the $V_{t2}$ plane (panel f), we observe two distinct populations: the incoming solar wind, $V_{t2}\sim$0\,km/s, and gyrating ions $V_{t2}\sim$500\,km/s, that are the reflected ions which have been accelerated by shock drift acceleration (SDA). This close correspondence suggests that the strongest Hall electric fields $E_n$ develop at times when the ramp and the reflection point are locally shifting downstream.

These observations establish three key properties of the shock: (1) pronounced non-stationarity evidenced by multiple phase-space holes, (2) strong Hall electric fields localized near the ion reflection point, and (3) a large, time-dependent reflected-ion population. In the following section, we compare these observations with two-dimensional hybrid simulations to determine the physical origin of these non-stationarity features.

\section{Simulation} 
We perform a two-dimensional hybrid particle-in-cell simulation using the HYPSI code \cite{Burgess2007,Trotta2020,Trotta2023turb,Vuorinen2026}, based on the current advance method and cyclic leapfrog (CAM-CL) algorithm \cite{Matthews1994}. In the model, protons are treated kinetically as macroparticles, while electrons are described as a massless, charge-neutralizing adiabatic fluid with $\gamma_e=5/3$. The shock is initiated with the injection method \cite{Quest1985}, where a super-Alfv\'enic plasma flow is continuously injected from the left boundary, reflected at the right boundary, and forms a shock propagating in the $-x$ direction, with periodic boundaries in $y$. The simulation is performed in a normal incidence frame (NIF) where the shock downstream has zero mean speed.

Quantities are normalized to their unperturbed, upstream values, with lengths in ion inertial lengths $d_i$, times in inverse cyclotron frequencies $\Omega_{ci}^{-1}$, velocities in Alfv\'en speed $v_A$, and fields and densities in $B_0$ and $n_0$, respectively.

In the simulation presented here, the upstream flow speed is $V_\mathrm{in}=\,4.5 \,v_A$, giving  $M_A\simeq6$ (similar to the MMS event), with Maxwellian upstream ions and $\beta_i=\beta_e=1$. The simulation domain is $50 \times 50,d_i^2$ with $\Delta x=\Delta y=0.2 d_i$, 500 particles per cell upstream, and is evolved to $20,\Omega_{ci}^{-1}$ using an ion time step $\Delta t=0.01,\Omega_{ci}^{-1}$. Thus, we have sufficient spatial resolution to resolve sub-$d_i$ structures associated with the Hall electric field, and enough particles per cell to keep statistical noise at a reasonable level. A background magnetic field $B_0$ is set up along the $\hat{z}$ direction and is therefore out of the simulation plane. Such a configuration disables the AIC instability, which requires an in-plane magnetic field component, so no ripples are expected to develop in the simulation. Meanwhile, the ion motion in the B-perp plane is fully resolved, which is essential for reformation.

\begin{figure}[!h]
\center
\includegraphics[width=1\textwidth]{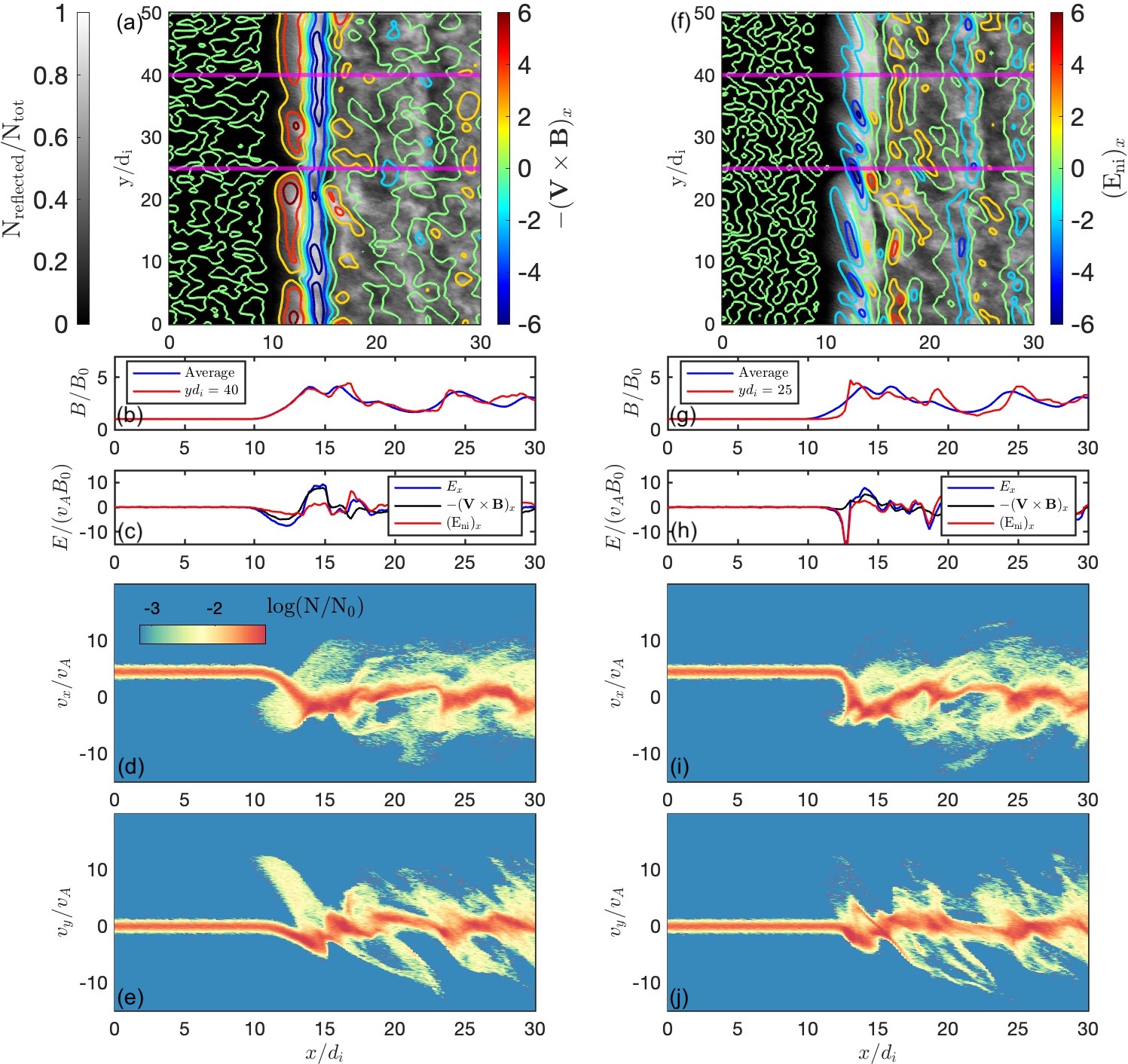}
\caption{Shock structure in 2D hybrid simulation. Top panels: fraction of reflected ions (black-and-white background) with contours of convection electric field $-(\mathbf{V_i} \times \mathbf{B})_x$ (panel a) and $x$-component (normal) of the non-ideal electric field $E_{ni}$ (panel f). 
The panels below show cuts (pink lines in panels a, f) at $y=40$ through a dense reflected-ion foot (left) and at $y=25$ through an almost foot-free shock transition (right). Panels (b,g): magnetic field magnitude (red) and the average over the box (blue),  panels (c, h): normal electric field, panels (d, i) $V_x-x$ reduced 1D ion VDF, panels (e, j) $V_y-x$ reduced 1D ion VDF.}
\label{fig:sim_cuts}
\end{figure}

To identify the physical origin of the observed shock non-stationarity, we examine the spatial structure of the simulated shock at $T\Omega_{ci}=15$. The shock front exhibits pronounced variations along its surface, with large changes in the reflected-ion population, magnetic profile, and electric field. To quantify the reflected ions, we compute their density, $N_{\rm reflected}$, by integrating the ion velocity distribution for $V_y>0$, which better separates the reflected population than using $V_x<0$. The resulting map is shown in the top panels of Figure~\ref{fig:sim_cuts}.

The reflected-ion population is highly structured along the shock front. In some regions, a well-developed foot contains reflected ions with densities approaching the upstream solar wind density, whereas in neighboring regions the foot is nearly absent. The structure of $(\mathbf{V}\times\mathbf{B})_x$ closely follows the reflected-ion density, while the non-ideal electric field, $E_{ni}$, exhibits the opposite behavior: its strongest localized peaks occur where the reflected-ion foot is weakest. These spatial variations naturally produce large differences in the local shock structure.

To investigate these regimes, we select two cuts normal to the shock through representative regions (pink lines in Figure~\ref{fig:sim_cuts}).

The first cut ($y=40$) passes through a dense reflected-ion foot. The magnetic field profile is smooth and follows the average shock profile, with the transition from the upstream field to the overshoot extending over approximately $3\,d_i$. The normal electric field is largely balanced by the ion convection term, $E_x\approx-(\mathbf{V}\times\mathbf{B})_x$, so ion dynamics remain close to ideal throughout most of the foot. The ion phase space shows the reflected ions ($v_y>0$) forming the foot, while the incoming ions are decelerated and deflected toward negative $v_y$ before reaching the ramp. This results from momentum exchange between the two populations, with the reflected ions gaining energy through SDA.

The second cut ($y=25$, right panels of Figure~\ref{fig:sim_cuts}) intersects a region where the reflected-ion population is almost absent. The magnetic transition is much sharper than the average profile, and the normal electric field is almost entirely supported by the non-ideal term, $E_{ni}$. The incoming ions reach the ramp with little deceleration and are reflected locally before gyrating away, so the reflected ions populate neighboring regions rather than forming a local foot. This combination of a narrow transition, a weak reflected-ion population, and an intense non-ideal electric field closely resembles the MMS crossings with the strongest $E_n$ spikes.

Neighboring regions of the same shock can thus exhibit dramatically different local structures because the reflected-ion population varies spatially. A spacecraft traversing such a shock would alternately encounter (1) broad, foot-dominated transitions governed by the convection electric field and (2) narrow ramps with intense Hall electric fields, naturally reproducing the principal signatures in Figure~\ref{fig:mms}.

As the shock evolves on ion timescales, a direct comparison between spatial cuts through the simulation and MMS observations is not straightforward. Unlike an instantaneous cut through the simulation domain, the spacecraft traverses the shock slowly, thereby sampling a mixture of spatial structure and temporal evolution. To enable a more direct comparison with the observations, we construct a virtual spacecraft (VSC) that slowly moves through the simulation in the normal incidence frame using the measured shock speed.

\begin{figure}[!h]
 \center
\includegraphics[width=1\textwidth]{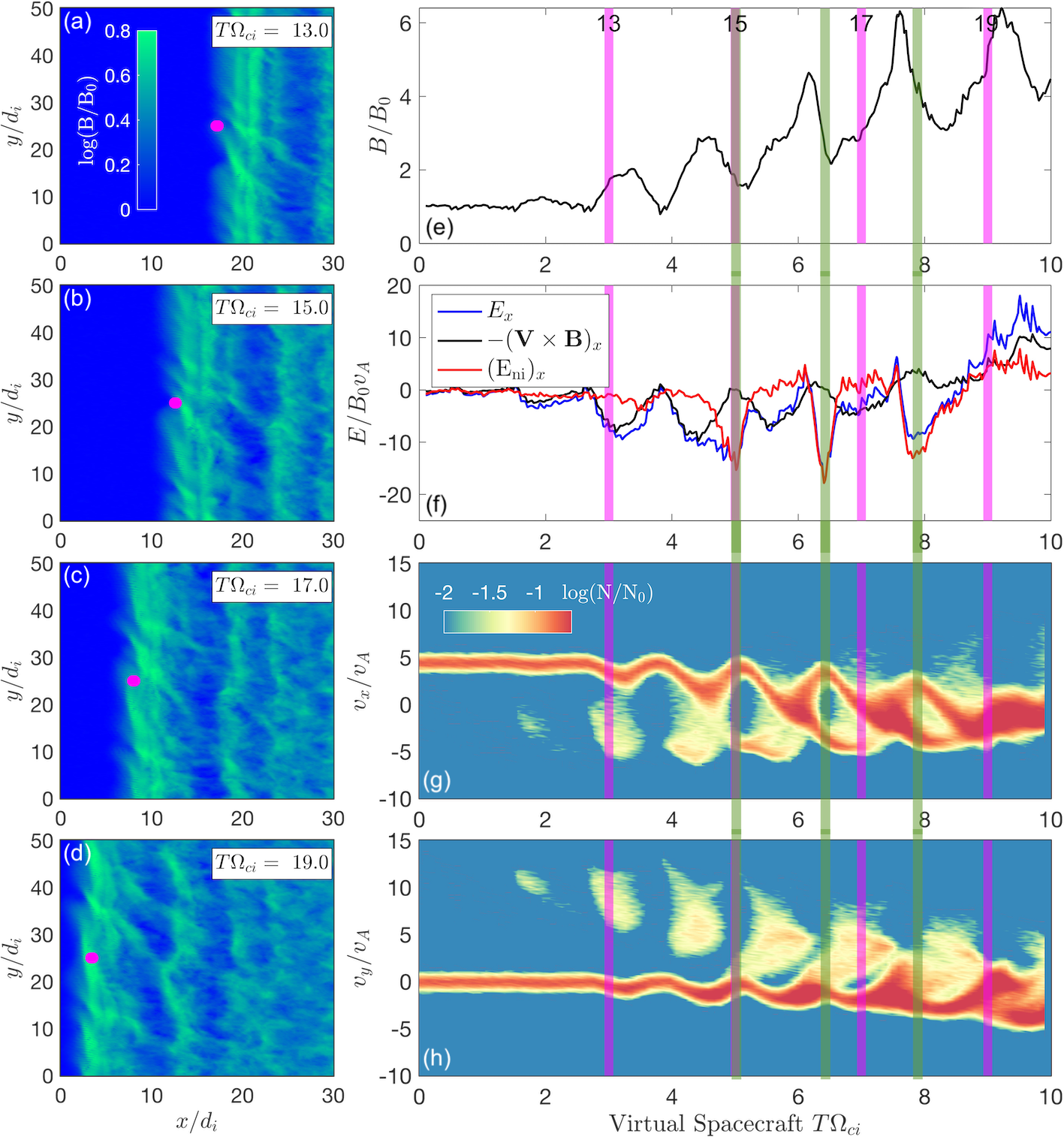}
\caption{Observations by the virtual spacecraft (VSC). Panels a-d show the time evolution of the magnetic field magnitude and the location of the virtual spacecraft (pink dot). Panels e-h show VSC observations in the same format as Figure~\ref{fig:sim_cuts}. The vertical pink stripes mark the times corresponding to panels a-d ($T \Omega_{ci}$=13,15,17,19). The vertical green stripes mark the peaks of the non-ideal electric field $E_{ni}$. }
\label{fig:sim_like_data}
\end{figure}

The left panels (a-d) of Figure~\ref{fig:sim_like_data} show two-dimensional maps of $|B|$ at successive times. The shock propagates in the negative $\mathbf{\hat{x}}$ direction (toward the upstream) and exhibits pronounced non-stationarity, with large variations in the overshoot magnetic field both along the shock front ($\mathbf{\hat{y}}$) and over ion-gyro timescales. These structures propagate toward positive $\mathbf{\hat{y}}$, i.e., in the direction of the reflected ions within the foot, indicating that the reflected-ion dynamics control the evolution of the shock surface. 
We interpret these propagating structures as the two-dimensional imprint of the reflected-ion reformation cycle discussed later.

The VSC, indicated by the pink dot, remains nearly stationary relative to the downstream plasma as the shock gradually overtakes it. The selected interval begins with the VSC upstream of the shock and ends after it has crossed the overshoot, closely reproducing the geometry of the MMS observations.

The right panels of Figure~\ref{fig:sim_like_data} show the measurements obtained by the virtual spacecraft (VSC) during its shock crossing. The simulation coordinate system $(\mathbf{\hat{x}},\mathbf{\hat{y}},\mathbf{\hat{z}})$ corresponds directly to the $(\mathbf{\hat{n}},\mathbf{\hat{t}_2},\mathbf{\hat{t}_1})$ system used for the MMS observations, enabling a direct comparison between the two data sets.

The VSC reproduces several of the principal observational signatures identified by MMS. The reduced ion velocity distributions exhibit multiple ion phase-space holes in the $v_x$ direction together with a reflected-ion population in the foot at $v_y>0$. The magnetic field displays several distinct compressions, each accompanied by a local decrease in the incident-ion velocity $v_x$, indicating repeated encounters with different parts of the evolving shock transition. Similarly, localized enhancements of the negative normal electric field are found at these compressions and are largely supported by the non-ideal electric field, $E_x\simeq E_{ni}$. What is most important is that the $E_{ni}$ spikes are observed at the times when $B$ is decreasing, thus when the ramp is locally retreating downstream, which is similar to the MMS observations.

These repeated structures correspond to partial crossings of the shock ramp as the structured shock surface sweeps over the virtual spacecraft. Only the final structure represents a complete transition into the downstream, where the incident ions have been fully thermalized ($v_x\approx0$) and the normal electric field changes sign ($E_x>0$). 
The close correspondence between the VSC and MMS observations demonstrates that the observed magnetic structures, ion phase-space holes, and localized Hall electric fields are naturally explained by repeated encounters with a structured shock front, where the structure is produced by shock reformation.

The simulation also differs from MMS observations in several important ways. First, because the simulation is two-dimensional, with the upstream magnetic field perpendicular to the simulation plane, it does not reproduce the variations in the normal magnetic field component, $B_n$, characteristic of ripples. Second, the morphology of the ion phase-space holes differs substantially. In the MMS observations, the holes are nearly symmetric and remain connected in phase space, whereas in the simulation, they are skewed and appear disconnected. These differences indicate 
that additional physics associated with 3D shock structure plays an important role.

\section{Discussion} 

In the simulation, the shock front is spatially structured, and at each location the shock evolves through a reformation-like reflection cycle. To interpret this structure, we consider the dynamics of the reflected ions. The normal component of the ion equation of motion is
\begin{equation}
\frac{\mathrm{d}}{\mathrm{dt}} V_n = \frac{e}{m_i} E_n - \frac{e}{m_i} V_{t2}B.
\label{eq:ion_vn}
\end{equation}

We consider two representative locations along the shock front (cuts in Figure~\ref{fig:sim_cuts}): a thin shock without a reflected-ion foot and a thick shock with a dense foot.

In the thin shock, the transition is a narrow, steep ramp. Incoming ions with $V_{t2}=0$ encounter a strong non-ideal electric field, see Eq.~\ref{eq:gol},
\begin{equation}
E_n \approx E_{ni} = \frac{1}{n_e e} (\mathbf{J} \times \mathbf{B})_n 
- \frac{1}{n_e e} \frac{\partial }{\partial \mathrm{n}} p_e =  - \frac{1}{n_e} \frac{\partial}{\partial \mathrm{n}} \left( \frac{B^2}{2\mu_o} + p_e \right),
\label{eq:E_n_thin}
\end{equation}
and the associated cross-shock potential specularly reflects a significant fraction of them (Figure~\ref{fig:sim_cuts}h-j). Such strong localized reflection is incompatible with a stationary shock.

The reflected ions gyrate upstream to form the foot, where the convection electric field accelerates them along positive $\hat{\mathbf{t}}_2$. Through the self-consistent fields of the reflected population, the incoming ions acquire a negative $V_{t2}$ (Figure~\ref{fig:sim_cuts}d-e). The larger the reflected population, the stronger the tangential deflection and the deceleration of the incoming ions through the $V_{t2}B$ term in Eq.~\ref{eq:ion_vn}.

The reflected ions also weaken the Hall field. They increase the local density and, because their current enhances the field within the foot, reduce the magnetic-field gradient across the ramp. Both effects reduce the Hall term in Eq.~\ref{eq:E_n_thin} and the cross-shock potential. Newly arriving ions are therefore reflected less efficiently and, having already been decelerated and deflected in the foot, acquire a much smaller upstream excursion than ions reflected from a thin ramp. The foot is no longer replenished and decays.

As the foot decays, the ramp steepens, restoring the Hall field and the cross-shock potential, and a new reflection cycle begins. The Hall field and the reflected-ion population thus form a self-regulating feedback cycle: the Hall field generates reflected ions, which in turn weaken the Hall field.

A connection between ion foot and electric-field structure has been demonstrated in one-dimensional simulations \cite{Lee2005POP,Lee2005AG,Yang2009POP,Yang2009JGR,Hao2023ApJ}. We extend this picture to a 2D, time-dependent shock, where reflected ions return at locations displaced from their reflection points. The resulting motion of reflected-ion bunches along the front is associated with the propagation of the magnetic structures and lets the local reflection cycle, Hall field, overshoot, and foot thickness evolve coherently along the shock.

\section{Conclusions}

We investigated a non-stationary, nearly perpendicular supercritical shock observed by MMS using 2D hybrid simulation. We performed spatiotemporal analysis of simulation data, which lets us track the shock evolution and link it directly to the signatures sampled by a virtual spacecraft. 
The simulation is in the plane perpendicular to the upstream magnetic field, which allows reformation but prevents the development of the AIC instability and associated ripples. 


The simulation establishes a direct link between the spatio-temporal evolution of the ion reflection, foot thickness, and the electric field in a two-dimensional system. Observations and simulation show that the shock transition develops neighboring regions with two distinct local states. In the first \emph{thin} state, the shock has a narrow ramp with little or no upstream foot. The normal electric field is dominated by the Hall term, producing a large cross-shock potential capable of specularly reflecting a substantial fraction of the incoming ions. In the second \emph{thick} state, the reflected ions have propagated upstream to form a dense foot extending over approximately one ion gyroradius. In this regime, $E_n$ is largely balanced by the ion convection term associated with the reflected-ion population, and the incoming ions are gradually decelerated and deflected within such a foot, so they interact with a shock in a magnetized way, in contrast to the non-magnetized interaction for ions hitting the thin ramp.

These two states are linked through a self-regulating feedback cycle. Strong Hall-field reflection produces the reflected-ion foot, while the foot subsequently weakens the Hall electric field, suppressing further reflection. As the reflected ions are convected away and transmitted downstream, the foot decays, the Hall electric field is restored, and a new reflection cycle begins. The result is a highly structured, non-stationary shock in which neighboring regions simultaneously occupy different phases of the reflection cycle.

The virtual-spacecraft analysis shows that this evolving shock naturally reproduces many of the qualitative features observed by MMS, including multiple magnetic compressions, localized Hall electric fields, ion phase-space holes, and intermittent reflected-ion populations. These results show that reformation physics can account for several principal signatures of the observed non-stationarity and link them to the self-regulating feedback between Hall-field ion reflection and the evolving reflected-ion foot.

At the same time, essential observational features are not reproduced by the simulation, including variations in the normal component of the magnetic field and the symmetric structure of the ion phase-space holes. These discrepancies indicate that reformation alone, as represented in the present two-dimensional geometry, does not capture the full observed shock structure. 
A fully three-dimensional treatment that includes both reformation and rippling physics is therefore needed to assess their respective contributions and reproduce the detailed morphology of the observed shock.

%
%

\section*{Conflict of Interest}
The authors declare no conflicts of interest relevant to this study.

\section*{Availability Statement} 
The Magnetospheric Multiscale (MMS) data used in this study are available at \citeA{mms4fgmbrst,mms4fpidisdistbrst,mms4fpidismomsbrst,mms4sdpEbrst}.
Data analysis was performed using the IRFU-Matlab analysis package \cite{khotyaintsev_2024_11550091}. The simulation data are available from \citeA{Trotta2026_simulation}.

\acknowledgments 
We thank the MMS team and instrument PIs for data access and support. This work was supported by the Knut and Alice Wallenberg Foundation (Dnr. 2022.0087) and the Swedish National Space Agency (SNSA). YK and DT acknowledge support from ESA through the Science Faculty - Funding reference ESA-SCI-E-LE-172.

%
%


%
%
%
%
%

\end{document}